\documentclass[sigconf]{acmart}

\usepackage{enumitem}


\AtBeginDocument{%
  }

\copyrightyear{2026}
\acmYear{2026}
\setcopyright{cc}
\setcctype{by}
\acmConference[CIKM '26]{Proceedings of the 35th ACM International Conference on Information and Knowledge Management}{November 07--11, 2026}{Rome, Italy}
\acmBooktitle{Proceedings of the 35th ACM International Conference on Information and Knowledge Management (CIKM '26), November 07--11, 2026, Rome, Italy}
\acmDOI{10.1145/3799682.3840910}
\acmISBN{979-8-4007-2539-5/2026/11}

\begin{document}

\title{GrOIL: Graph-Grounded Domain Ontology Induction with Constrained LLM Mediation}

\author{Maruf Ahmed Mridul}
\affiliation{%
    \institution{Rensselaer Polytechnic Institute}
    \city{Troy}
    \state{New York}
    \country{USA}
    }
\email{mridum@rpi.edu}
\orcid{0009-0003-7501-4714}

\author{Abid Talukder}
\affiliation{%
    \institution{Rensselaer Polytechnic Institute}
    \city{Troy}
    \state{New York}
    \country{USA}
    }
\email{taluka@rpi.edu}
\orcid{0009-0009-9544-8994}

\author{Oshani Seneviratne}
\affiliation{%
	\institution{Rensselaer Polytechnic Institute}
	\city{Troy}
	\state{New York}
	\country{USA}
}
\email{senevo@rpi.edu}
\orcid{0000-0001-8518-917X}

\renewcommand{\shortauthors}{Mridul, Talukder and Seneviratne}

\begin{abstract}
Constructing formal ontologies from domain documents requires simultaneously enforcing corpus grounding, vocabulary consistency,  axiom-level expressivity, and end-to-end provenance, a combination no existing automatic system delivers. We present a seven-stage  graph-grounded pipeline that converts domain documents into a complete, auditable Web Ontology Language (OWL) Terminological Box (TBox) without any unconstrained generation step. Documents are first encoded as Unified Discourse-Hypergraphs (UDH) capturing entity participation and discourse dependencies; subsequent stages transform  this graph evidence into a class hierarchy, typed object and datatype properties, and restriction axioms, with Large Language Model (LLM) usage restricted to narrow, graph-grounded mediation tasks. A paired Assertional Box (ABox) population procedure grounds named individuals in the induced TBox, enabling SPARQL-based functional evaluation. Every emitted term carries a full decision chain from raw source passages through each pipeline stage, making the TBox directly auditable and suitable for targeted human refinement. Evaluated on the life insurance domain using two established benchmarks and a new 100-contract corpus spanning ten product types, our pipeline achieves strong results across all evaluation dimensions, outperforming direct and multi-agent LLM baselines on competency-question (CQ) coverage (\textbf{0.85} vs. 0.63 and 0.62 on one term-life contract; \textbf{0.77} vs. 0.40 and 0.44 on another contract), while also attaining high keyphrase coverage comparable to a manually-constructed reference ontology and strong performance on structured gap-and-overlap reasoning, all without any manual TBox engineering. Ontology growth analysis provides evidence consistent with vocabulary saturation at scale, demonstrating that the pipeline produces stable, reusable domain representations from large document corpora.
\end{abstract}





\maketitle

\vspace{-6pt}
\section{Introduction}

Ontologies remain a central mechanism for turning domain knowledge into reusable, inspectable, and executable representations. In high-stakes domains such as insurance, legal compliance, healthcare, and finance, they provide more than semantic organization: they make policies, obligations, exclusions, eligibility conditions, and decision rules available for formal querying and downstream reasoning. However, constructing such ontologies from domain documents remains difficult. Expert-built ontologies are accurate but costly to produce and maintain, while automatic ontology learning systems have 
historically struggled to induce coherent and rich axiomatic structure ~\cite{du2024short,lourdusamy2019survey}.

Automatic ontology construction methods, ranging from lexico-syntactic pattern mining  and embedding-based clustering to recent LLM-based approaches, have significantly  advanced the field, yet they share a persistent gap: systems either produce shallow outputs grounded in corpus evidence, or richer schema-level abstractions that lack 
grounding, consistency, and provenance~\cite{du2024short, lourdusamy2019survey, bakker2024ontology, babaei2023llms4ol}. The root cause is architectural: existing pipelines either delegate too much to the LLM, inviting hallucination, redundancy, and ungrounded term invention, or too little, forfeiting the semantic judgment that only a language model can supply. No existing system jointly enforces corpus grounding, vocabulary-level control, axiom-level expressivity, and end-to-end provenance within the same pipeline.

We argue that ontology induction should be treated as a grounded decision pipeline in  which the document corpus first supplies a structured evidence graph, and the LLM is  invoked only for bounded semantic decisions that resist deterministic resolution. Rather than asking the model to generate an ontology freely, every LLM call is restricted to a local modeling decision over graph-derived evidence and a closed output vocabulary, enforced at generation time via closed-vocabulary prompting, covering tasks such as naming clustered candidates, resolving ambiguous domain and range, and placing classes in an underspecified hierarchy. This design makes the LLM's role precise and auditable: it fills bounded gaps in structured reasoning, never authors structure freely.

We present a seven-stage pipeline that realizes this design, converting  documents into UDH using Talukder et al.'s ~\cite{udh2026} method and transforming the resulting graph  evidence into a complete, traceable OWL TBox. Every emitted term carries a complete decision chain, making the induced TBox auditable and suitable for targeted human refinement. We further extend the pipeline with an ABox population procedure enabling SPARQL-based functional  evaluation over populated ontologies.

We evaluate the system in the life insurance domain using two benchmarks ~\cite{mridul2026benchmark, talukder2026towards}, and a new 100-contract corpus spanning diverse product types. Insurance contracts are a difficult testbed: state-of-the-art models reading contracts directly still make systematic errors on coverage determination tasks, confirming that surface language understanding alone is insufficient and that structured, provenance-grounded representations carry genuine value ~\cite{mridul2026benchmark}. Our induced ontology achieves high scores on this benchmark, matching the strongest LLM baseline, without any manual TBox engineering, while additionally providing deterministic reproducibility and structured evidence trails. The contributions of this work are:

\begin{itemize}[topsep=2pt, itemsep=0pt, parsep=0pt, leftmargin=0.75em]
    \item \textbf{End-to-end TBox and ABox induction.} A graph-grounded pipeline that induces a complete OWL TBox and populated ABoxes from domain documents, with every class, property, restriction, and individual carrying full provenance annotations traceable to source passages.

    \item \textbf{Constrained LLM mediation.} The pipeline assigns LLMs narrow, well-defined roles within a graph-grounded ontology induction process. LLMs only mediate localized decisions, such as cluster naming, domain and range resolution, and class placement, while closed-vocabulary prompting constrains term-selection steps.

    \item \textbf{An extended 100-contract life insurance dataset.} We expand the 10-contract benchmark of Mridul et al.~\cite{mridul2026benchmark} to 100 synthetic contracts across diverse product types, enabling scalability and ontology-growth analysis beyond small benchmarks.

    \item \textbf{Comprehensive ontology evaluation.} We evaluate structural quality, CQ coverage, keyphrase coverage, ontology growth, and executable gap and overlap analysis, showing strong functional performance without manual TBox engineering.
\end{itemize}

The generated resources and our source code are available at \url{https://github.com/brains-group/GrOIL}, with the accompanying README providing detailed instructions for reproducibility.

\vspace{-0.7em}
\section{Related Work}
\label{sec:related-work}
Ontology engineering has long been recognized as a labor-intensive, expert-driven discipline, with a persistent \emph{knowledge acquisition bottleneck} at every phase from initial modeling through long-term maintenance~\cite{noy2001ontology,studer1998knowledge,gruber1993translation,tudorache2020ontology}. The ontology learning field emerged to attack this bottleneck by automating extraction from text~\cite{maedche2001ontology,cimiano2006ontology}. Early work established lexico-syntactic patterns for taxonomy induction~\cite{hearst1992automatic}: an approach that remains competitive on large corpora~\cite{roller2018hearst}, and the \emph{ontology learning layer cake} framing organized pipelines from term extraction through concept formation, relation extraction, and axiom induction~\cite{biemann2005ontology,asim2018survey}. Systems such as OntoLearn demonstrated that domain corpora can yield term hierarchies without manual seeding~\cite{navigli2004learning, velardi2013ontolearn}, yet scalable end-to-end solutions have remained elusive: relation extraction and axiom induction are consistently the hardest open problems, and recent reviews confirm that modern methods still assemble isolated components rather than delivering unified frameworks~\cite{du2024short,lourdusamy2019survey}.

Pre-trained language models transformed the upstream information extraction task that any ontology learning pipeline depends on~\cite{devlin2019bert}, and dense sentence representations enabled embedding-based deduplication of surface-variant concepts at scale~\cite{reimers2019sentence}: a technique our pipeline employs directly. LLMs have since attracted attention as ontology engineering assistants, but the landscape remains fragmented. LLMs4OL established that LLMs handle isolated sub-tasks such as term typing but degrade on more complex relational extraction tasks~\cite{babaei2023llms4ol}; OLLM showed that end-to-end taxonomic backbone construction is feasible but that domain adaptation requires fine-tuning on domain-specific examples~\cite{lo2024end}. Lippolis et al.\ show that careful prompting can outperform novice modelers on small benchmarks, yet without systematic corpus grounding~\cite{lippolis2025ontology}; Saeedizade and Blomqvist enumerate OWL modeling options from requirements but delegate selection to humans~\cite{saeedizade2024navigating}; Kommineni et al.\ pair open-source LLMs with an LLM judge but find quality varies sharply with model size~\cite{kommineni2024human}. Bakker et al.\ document persistent failures including hallucination, incomplete coverage, and unstable property typing~\cite{bakker2024ontology}, while Norouzi et al.\ demonstrate that ontology population is reliable when the LLM is given a closed vocabulary~\cite{norouzi2025ontology}---a principle we extend to domain and range assignment and restriction induction. Domain-oriented LLM efforts follow the same pattern: Chowdhury et al.\ and Abolhasani and Pan apply guided prompting to abstract cultural concepts and technical texts, respectively~\cite{chowdhury2025automated,abolhasani2024leveraging}, but both are corpus-specific. Talukder et al.\ directly precede our work with a multi-agent LLM baseline for insurance ontology generation that our pipeline substantially surpasses~\cite{talukder2026towards}.

Enforcing valid ontology term output requires constrained decoding; structured generation has become standard for reliable LLM deployment in machine-facing pipelines~\cite{willard2023efficient, beurer2024guiding, geng2025generating}.
At the broader LLM--KG integration level, Pan et al.\ map three integration regimes (KG-enhanced LLMs, LLM-augmented KGs, synergized systems)~\cite{pan2024unifying}, and Zhu et al.\ empirically confirm that large models serve better as inference assistants than few-shot extractors~\cite{zhu2024llms}; Bian et al.\ survey the resulting schema-based versus schema-free construction paradigms~\cite{bian2025llm}. Document-level relation extraction has benefited from graph-structured representations that capture multi-hop reasoning paths beyond sentence boundaries~\cite{yao2019docred}, motivating our graph formalism. Surface-variant deduplication via agglomerative clustering on dense embeddings~\cite{mullner2011modern} is a prerequisite for any induction system that must avoid TBox redundancy.

Domain ontologies for insurance and legal text confirm that formal ontologies enable 
executable SPARQL-based reasoning that LLM inference cannot replicate~\cite{ahaggach2023information,
charalambous2022analyzing,el2017towards,naqvi12023ontology}, and restriction axioms such as SWRL 
rules have clear downstream utility in high-stakes settings~\cite{amith2022expressing,horrocks2004swrl}. 
Competency questions and LLM-as-judge panels provide the standard evaluation frame~\cite{araujo2016data,
bezerra2013evaluating,liu2023g,zheng2023judging}, complemented by keyphrase coverage~\cite{mridul2026benchmark}, 
all of which we adopt.
Across this body of work a consistent gap emerges: classical methods are corpus-grounded but shallow; LLM-based methods produce syntactically valid OWL but lack global consistency and provenance; domain-specific efforts require substantial manual TBox engineering. Our pipeline sits at the intersection: grounding every LLM call in corpus-derived graph evidence, enforcing outputs against a closed ontology vocabulary, and materializing end-to-end provenance annotations on every emitted term.

\section{Methodology}
\label{sec:methodology}

Given a corpus $\mathcal{D} = \{d_1, d_2, \ldots, d_n\}$ of domain documents, the task is to produce a formal OWL TBox $\mathcal{T} = (\mathcal{C}, \mathcal{P}, \mathcal{A})$, where $\mathcal{C}$ is a set of named classes organized in an \texttt{rdfs:subClassOf} hierarchy, $\mathcal{P}$ is a set of object and datatype properties with explicitly typed domain and range, and $\mathcal{A}$ is a set of TBox restriction patterns (existential and universal restrictions, disjointness assertions, and cardinality constraints), all grounded in evidence drawn from $\mathcal{D}$.

\vspace{-2pt}
\subsection{Input Preparation}
Each document $d \in \mathcal{D}$ is first converted into a structured graph representation before any ontological reasoning is performed. We adopt the UDH formalism ~\cite{udh2026}, which encodes document knowledge along two complementary axes: \textit{entity-centric participation}, capturing which entities co-occur within the same proposition and in what functional roles, and \textit{narrative connectivity}, capturing the logical and conditional dependencies between propositions across the document.

\begin{figure}[t]
    \vspace{-2pt}
    \centering
    \includegraphics[width=1\linewidth]{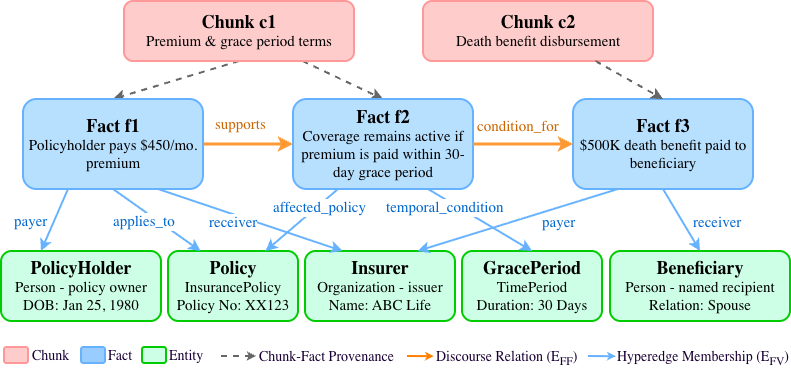}
    \caption{\normalfont \textbf{Input graph structure}: chunk nodes (provenance anchors), fact nodes (hyperedges over entities and discourse graph nodes), and entity nodes with typed roles and relational anchors.}
\label{fig:input-graph-structure}
\vspace{-5pt}
\end{figure}

\paragraph{\textbf{Graph Structure:}} A UDH for a document is defined as $\mathcal{G} = (C, F, V, E_{CF}, E_{FV}, E_{FF})$, as illustrated in Figure~\ref{fig:input-graph-structure}. \textit{Chunks} $C = \{c_1, \ldots, c_k\}$ are overlapping source-text windows used as provenance anchors. \textit{Facts} $F = \{f_1, \ldots, f_m\}$ are  atomic, self-contained propositions extracted from the text, each augmented with thematic topics, and a set of participating entities. Critically, facts occupy a dual structural role: they are simultaneously \textit{hyperedges} over the entities that participate in them (via $E_{FV}$) and \textit{nodes} in a directed discourse graph (via $E_{FF}$). \textit{Entities} $V = \{v_1, \ldots, v_p\}$ are richly typed fingerprints, each carrying a canonical name, an ontological type (e.g., \texttt{PERSON}, \texttt{ORGANIZATION}), a contextual micro-role grounding the entity's function within its passage, and a set of discriminating features such as policy numbers or relational attributes.

The two edge sets encode the two structural axes described above. The hyperedge-entity incidence relation $E_{FV} \subseteq F \times V \times \mathcal{L}$, where $\mathcal{L}$ is a vocabulary of edge labels, records the typed functional role of each entity in a fact (e.g., \textit{payer}, \textit{receiver}, \textit{affected\_policy}). The discourse edge relation $E_{FF} \subseteq F \times F \times \mathcal{L} \times [0,1]$ records directed, labeled narrative relationships between facts (e.g., \textit{supports}, \textit{condition\_for}), together with a confidence score. Together, these two edge sets make UDH graphs a rich substrate for ontology induction as evidenced in ~\autoref{fig:input-graph-structure}
: $E_{FV}$ provides direct, corpus-grounded evidence for property axioms, while $E_{FF}$ surfaces logical dependencies that would be invisible to entity or triple-based representations, a richness of structure that neither a flat chunk nor a binary triple can express.

\paragraph{\textbf{Document-to-Graph Conversion:}} We apply the full UDH construction pipeline ~\cite{udh2026} to each document independently, producing a collection of per-document graphs $\{\mathcal{G}_d\}_{d \in \mathcal{D}}$, each serialized as a typed GraphML file in which node records carry the full entity fingerprint metadata and edge records carry role labels, discourse relation types, and confidence scores. No raw document text is accessed beyond this point; all subsequent stages operate exclusively on the structured graph representation.

\paragraph{\textbf{Graph Cleaning:}}
Before ontology construction begins, each graph $\mathcal{G}_d$ is passed through a two-step cleaning pipeline applied to all edges $E_{FV} \cup E_{FF}$. 
\textbf{Step~1} removes null-like edges (such as \textit{none}, \textit{unrelated}, or \textit{independent}), discards low-confidence edges, and normalizes labels via a synonym map of a predefined set of high-frequency pre-defined canonical relation types. \textbf{Step~2} globally clusters surviving labels using agglomerative clustering~\cite{mullner2011modern} to resolve near-paraphrases not covered by the map, flagging suspicious nulls and rare labels for removal, and eliminating self-loops and duplicate triples.
The cleaned graphs  $\{\mathcal{G'}_d\}$ then serve as input to Stage~1 of the ontology construction pipeline.

\vspace{-2pt}
\subsection{Pipeline Overview}
\label{subsec:overview}
\begin{figure*}[ht]
    \centering
    \includegraphics[width=0.78\linewidth]{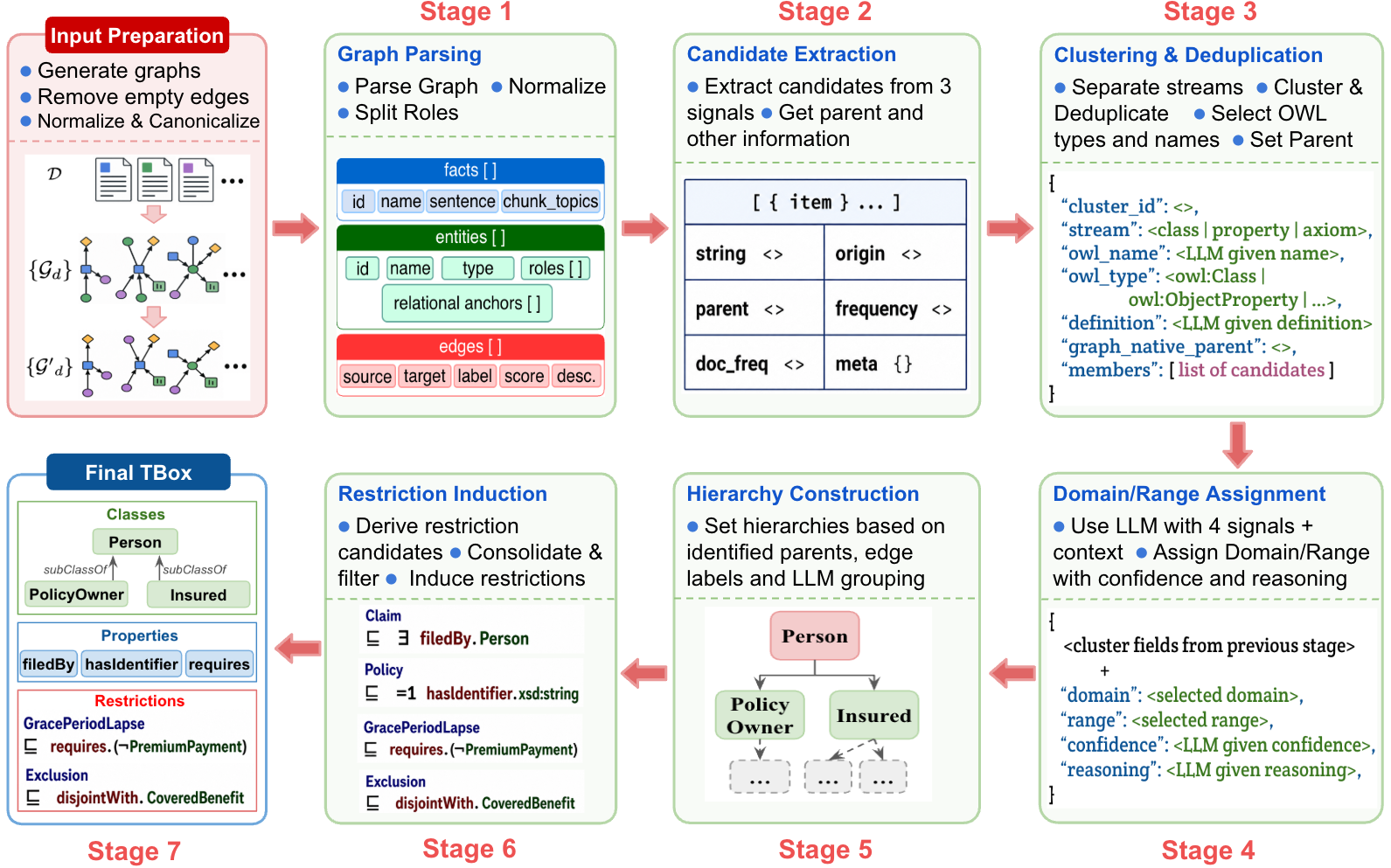}
    \caption{Overview of the proposed ontology construction pipeline.}
    \label{fig:pipeline}
    \vspace{-5pt}
\end{figure*}

Ontology construction from $\{\mathcal{G}'_d\}_{d \in \mathcal{D}}$ proceeds in seven stages, summarized in \autoref{fig:pipeline}. The pipeline prioritizes graph-derived, corpus-grounded evidence before invoking LLMs for bounded decisions, including naming, cluster formalization, ambiguous domain and range assignment, and underspecified hierarchy construction. This restricts the model's freedom, grounds its outputs in corpus evidence, and reduces reliance on large proprietary models, making small open-source LLMs sufficient.

\vspace{-2pt}

\subsection*{Stage 1: Parsing and Typed Pool Construction}
\label{subsec:stage1}
Stage 1 reads the cleaned GraphML files in configurable batches and produces, for each document, a \textbf{typed pool}: the partitioned sets of chunk, fact, and entity records, together with an edge bucket index mapping each bucket label to its edges. Each entity record carries five fields: an identifier, a name, a normalized type, a roles list, and a set of relational anchors. 
The \texttt{roles} field deserves attention: entities frequently occupy multiple functional positions encoded as compound role strings (e.g., \textsc{PERSON} : \textit{Policy Owner and Insured}). We split these along conjunctive delimiters, yielding atomic roles that each become independent class candidates in Stage~2, with co-occurring roles providing sibling signals for Stage~5.

The relational anchors (e.g., \textit{DOB: January 25, 1980; Age: 44; Underwriting Class: Standard Plus Non-Tobacco}) are preserved verbatim as structured metadata, providing a pool of data property candidates recoverable in later stages.

\vspace{-2pt}
\subsection*{Stage 2: Concept Candidate Extraction}
\label{subsec:stage2}

Stage~2 aggregates the typed pools across all documents in $\mathcal{D}$ and 
extracts a flat \textbf{candidate pool} $\Gamma$. Each candidate $\gamma \in 
\Gamma$ is a tuple:
\[
  \gamma = \bigl(\,\underbrace{s}_{\text{string}},\;
                  \underbrace{o}_{\text{origin}},\;
                  \underbrace{p}_{\text{parent}},\;
                  \underbrace{f}_{\text{freq}},\;
                  \underbrace{df}_{\text{doc\_freq}},\;
                  \underbrace{m}_{\text{meta}}\,\bigr)
\]
where $o \in \{\textsc{entity\_type},\, \textsc{entity\_role},\, 
\textsc{edge\_label},\, \textsc{fact\_name},\, \\\textsc{chunk\_topics}\}$ records 
the graph signal from which the candidate was extracted, $p$ encodes a 
structural subclass hypothesis for role-derived candidates (a role observed 
under type $t$ is hypothesized to denote a subclass of the class corresponding 
to $t$), and $m$ carries auxiliary evidence (primarily source and target type 
distributions for edge-label candidates). Candidates are extracted from the following signals.

\vspace{-0.5em}
\paragraph{\textbf{Entity Types and Roles:}}
For every entity in the corpus, one candidate is emitted for its normalized 
type and one for each atomic role in its roles list. The origin tag 
distinguishes these: \textsc{entity\_type} for type-derived candidates and 
\textsc{entity\_role} for role-derived ones. For the latter, the \texttt{parent} 
field records the entity type under which the role was observed, directly 
encoding the ontological hypothesis that the role denotes a subclass of the 
corresponding type class. An instance of \textit{policy owner} observed under a 
\textsc{person} entity encodes $\mathit{PolicyOwner} \sqsubseteq \mathit{Person}$ 
without consulting any external lexical resource. This structural parent link 
propagates to Stage~3, where it provides high-confidence subclass placement if 
observed consistently across the corpus.

\vspace{-0.5em}
\paragraph{\textbf{Edge Labels with Type Context:}}
For each edge in the corpus, its label is emitted as an \textsc{edge\_label} 
candidate. The metadata field records the empirical distributions of source and 
target entity types observed for that label across the corpus. A label such as 
\textit{has death benefit option}, consistently observed between \textsc{policy} 
source nodes and fact-node targets, carries a distributional prior exploited for 
domain and range assignment in Stage~4. Collecting these distributions at 
extraction time keeps subsequent stages independent of the raw graph files.

\vspace{-0.5em}
\paragraph{\textbf{Fact Names:}}
Fact node names are emitted as \textsc{fact\_name} candidates, surfacing 
propositional predicates (such as \textit{policy lapse after grace period} or 
\textit{premium payment frequency}) that are unreachable through entity or edge 
signals alone. Fact names also feed the axiom stream in Stage~3, where names 
containing modal or conditional language are treated as axiom candidates rather 
than plain property candidates.

\vspace{-0.5em}
\paragraph{\textbf{Chunk Topics:}}
Thematic topic strings associated with facts are emitted as 
\textsc{chunk\_topics} candidates. These are extracted from the 
\texttt{chunk\_topics} field stored on fact nodes, since Stage~1 records topics 
via facts rather than directly via chunk records, and serve as supplementary 
class candidates in Stage~3's class stream.

After candidate extraction, frequency filtering is applied per origin 
to ensure candidates are representative across the corpus rather than inflated 
by a single verbose document, controlling for distributional skew and improving 
generalizability.

\subsection*{Stage 3: Deduplication and Semantic Clustering}
\label{subsec:stage3}

The candidate pool $\Gamma$ contains considerable lexical redundancy: surface variants of the same underlying concept must be consolidated before OWL terms can be introduced. Stage 3 performs this via embedding-based agglomerative clustering followed by LLM-assisted normalization, operating independently across three candidate streams.

\vspace{-0.5em}
\paragraph{\textbf{Stream Separation:}} Candidates are partitioned into three streams. The \textbf{class stream} takes candidates of origin \textsc{entity\_type}, \textsc{entity\_role}, and \textsc{chunk\_topics}. The \textbf{property stream} takes \textsc{edge\_label} candidates and \textsc{fact\_name} candidates whose string does not match a restriction keyword pattern. The \textbf{restriction stream} takes \textsc{fact\_name} candidates whose underlying proposition requires a logical operator to express in OWL. Since surface wording alone is unreliable for this distinction, stream assignment for \textsc{fact\_name} candidates is delegated to an LLM, which classifies each using its source sentence as evidence. Streams are processed independently to prevent cross-stream conflation: a string that functions as both a class name and a relation predicate in different contexts is handled separately in each stream.

\begin{figure}[t]
    \centering
    \includegraphics[width=1\linewidth]{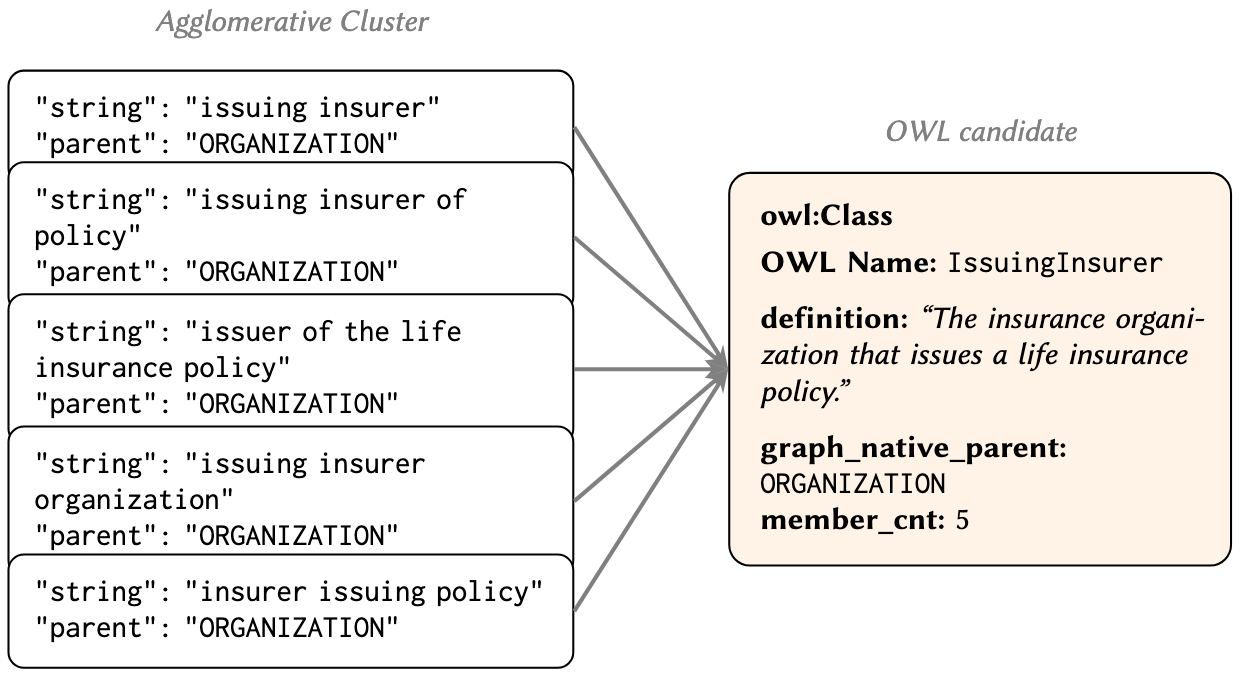}
    \caption{\normalfont\textbf{Example clustering outcome for a class-stream cluster}. Five surface variants, all observed under entity type \textsc{organization}, are merged by agglomerative clustering into a single \texttt{owl:Class} candidate. The graph-native parent is assigned here, because all variants carry the same parent information from the graph.}
\label{fig:cluster}
\end{figure}

\vspace{-0.5em}
\paragraph{\textbf{Embedding-Based Clustering:}} Within each stream, candidate strings are encoded using \texttt{all-mpnet-base-v2}\footnote{https://huggingface.co/sentence-transformers/all-mpnet-base-v2} with origin-aware prompt templates that inject domain context before embedding. An \textsc{entity\_role} candidate with parent $p$ is rendered as \textit{``insurance role of $p$: \{string\}''}, an \textsc{edge\_label} candidate as \textit{``relationship in insurance policy: \{string\}''}, and so on. The resulting normalized embeddings are clustered with \textbf{agglomerative clustering} ~\cite{mullner2011modern} (average linkage, cosine distance, similarity threshold). The threshold-based stopping criterion means no pre-specified $k$ is required; the number of clusters is determined entirely by the data geometry. Average-linkage cosine clustering is well-suited here because, after frequency filtering, the candidate strings within each stream are already semantically coherent: near-paraphrase variants (see ~\autoref{fig:cluster}) lie within a tight cosine neighborhood and merge naturally, while semantically distinct concepts remain separated.

\vspace{-0.5em}
\paragraph{\textbf{LLM-Assisted Normalization:}} For each cluster, the top-$k$ members by frequency are presented to the LLM, which assigns a canonical OWL name in PascalCase (for classes) or camelCase (for properties) and generates a concise natural-language definition. The LLM also assigns the type (\texttt{owl:ObjectProperty} or \texttt{owl:DatatypeProperty}) for properties and \texttt{RestrictionPattern} for the restrictions at this stage. After naming, parent assignment for class-stream clusters proceeds by structural majority vote. For any cluster that contains at least one \textsc{entity\_role} member, the \texttt{parent} field of each such member (set in Stage~2, Section~\ref{subsec:stage2}) is collected, and the most frequent parent type is recorded as \texttt{graph\_native\_parent} at confidence level \textit{high}. For clusters with no role members, or where role members lack consistent parent information, the parent field is left unresolved here and delegated to Stage~5.

\subsection*{Stage 4: Domain and Range Assignment}
\label{subsec:stage4}

This stage annotates each \texttt{owl:ObjectProperty} and \texttt{owl:DatatypeProperty} cluster from the property and restriction streams with a typed domain and range. The key design choice is to treat this as a multi-signal decision problem: four complementary graph signals are computed from the corpus and presented collectively to an LLM, which synthesizes them into a final assignment under a hard vocabulary constraint enforced at generation time.

\textbf{S1 (edge topology)} aggregates the empirical source and target node-type
distributions over all edges whose label is non-null and loosely matches any member
of the property cluster. The mode source type is reported as a domain candidate and
the mode target type as a range candidate. A minimum edge count threshold filters out property clusters with insufficient corpus evidence
before any nomination is made.

\textbf{S2 (edge direction)} specifically examines the \textsc{entity\_fact} and
\textsc{fact\_entity} edge buckets. An entity appearing as the source of an
\textsc{entity\_fact} edge with a matching label is a domain candidate; an entity
appearing as the target of a \textsc{fact\_entity} edge is a range candidate. This
signal attends to the directed argument structure encoded in the graph topology
rather than the global type distribution.

\textbf{S3 (raw sentences)} and S4 \textbf{(chunk topics)} supply natural-language evidence directly to the LLM: S3 passes up to $k_{s}$ fact-node sentences verbatim, while S4 passes the top $k_{t}$ thematic topic strings ranked by corpus frequency. Both are used as-is without further parsing.

\begin{figure}[t]
    \centering
    \includegraphics[width=1\linewidth]{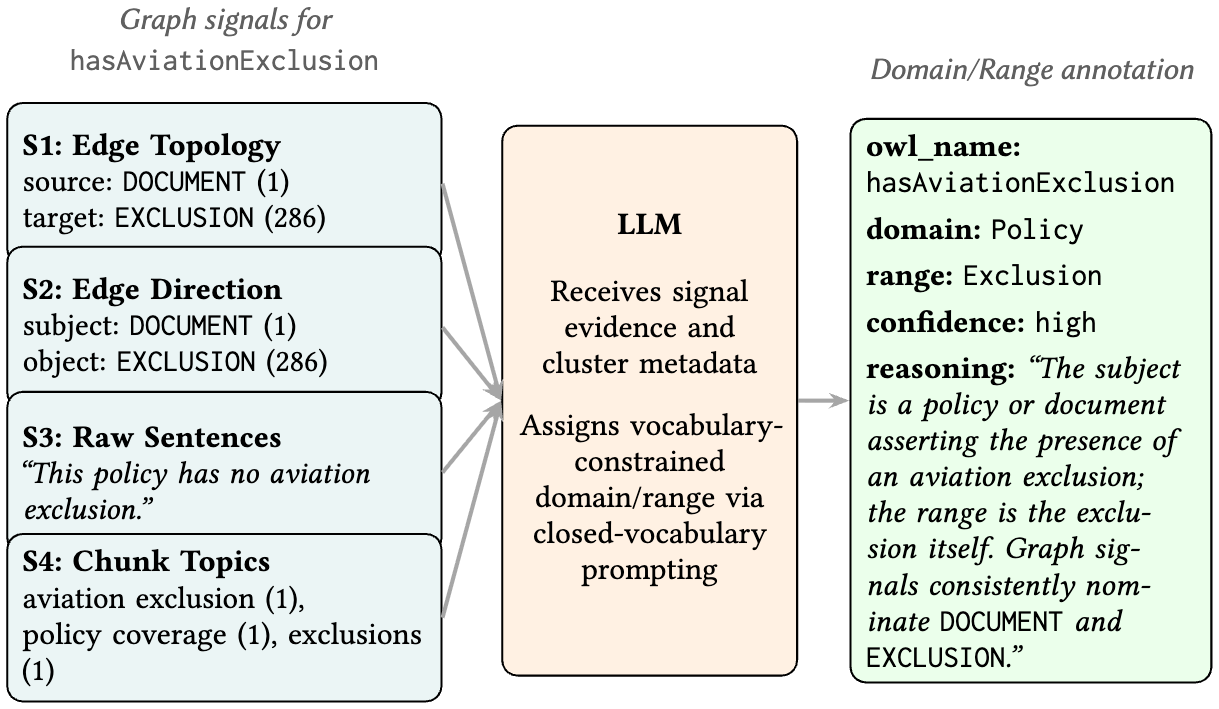}
    \caption{\normalfont\textbf{Domain and range assignment for}
    \texttt{hasAviationExclusion}. Four graph signals guide the decision:
    S1 and S2 nominate \texttt{DOCUMENT} as domain and \texttt{EXCLUSION}
    as range, while S3 and S4 provide supporting natural-language context.
    The LLM synthesizes these signals under closed-vocabulary prompting,
    producing a high-confidence assignment with an explanatory reasoning.}
    \vspace{-6pt}
    \label{fig:domrange}
\end{figure}

\paragraph{\textbf{LLM Assignment with closed-vocabulary prompting: }} All four signals, together with the cluster's OWL name, definition, and top member strings, are presented to the LLM in a structured prompt. The LLM selects domain and range from a closed vocabulary: the set of OWL class names produced in Stage~3 for object properties, and a fixed set of XSD datatypes for datatype properties. Adherence to this vocabulary is enforced at the generation level via closed-vocabulary prompting, so that the output is always a valid ontology term regardless of signal quality. The LLM additionally returns a confidence rating (high, medium, or low) and a brief reasoning string. Low-confidence assignments are flagged automatically for human review and are not committed to the TBox until inspected.

As illustrated in ~\autoref{fig:domrange}, the signals provide quantitative, corpus-grounded evidence; the LLM makes the final decision, using domain knowledge to resolve cases where signals conflict, are absent, or point in the wrong direction. The last case is a known extraction artifact: the graph extraction model does not always direct edges consistently, so S1 and S2 may nominate swapped domain and range values. The LLM is explicitly instructed to treat signals as hints and defer to insurance semantics, allowing it to detect and correct these inversions.

\vspace{-2pt}
\subsection*{Stage 5: Hierarchy Construction}
\label{subsec:stage5}

Stage 5 organizes the class clusters produced in Stage~3 into a formal
\texttt{rdfs:subClassOf} hierarchy.
Because Stage~3 classes are extracted bottom-up from corpus evidence, they
lack organizing abstractions at the top.
This stage addresses this via two phases.

\paragraph{\textbf{Phase 1: Abstract Class Discovery:}} A dedicated LLM call examines the full inventory of Stage~3 class names and
definitions and proposes a small set of abstract top-level classes that are domain-appropriate but do not appear in the corpus directly.
This is necessary because bottom-up extraction surfaces only the concepts explicitly named in documents.
A corpus of insurance contracts may yield \texttt{SuicideExclusion} and
\texttt{WarExclusion} as distinct
classes, each grounded in specific policy language, while the organizing abstraction \texttt{Exclusion} was never named as such in any document.
The abstract class discovery step recovers these missing abstractions.
Proposed classes are constrained to be mutually exclusive in intent,
preventing competing proposals from being nominated for the same children.

{
\setlength{\abovecaptionskip}{2pt}
\setlength{\belowcaptionskip}{-4pt}
\begin{figure}[t]
    \centering
    \includegraphics[width=1\linewidth]{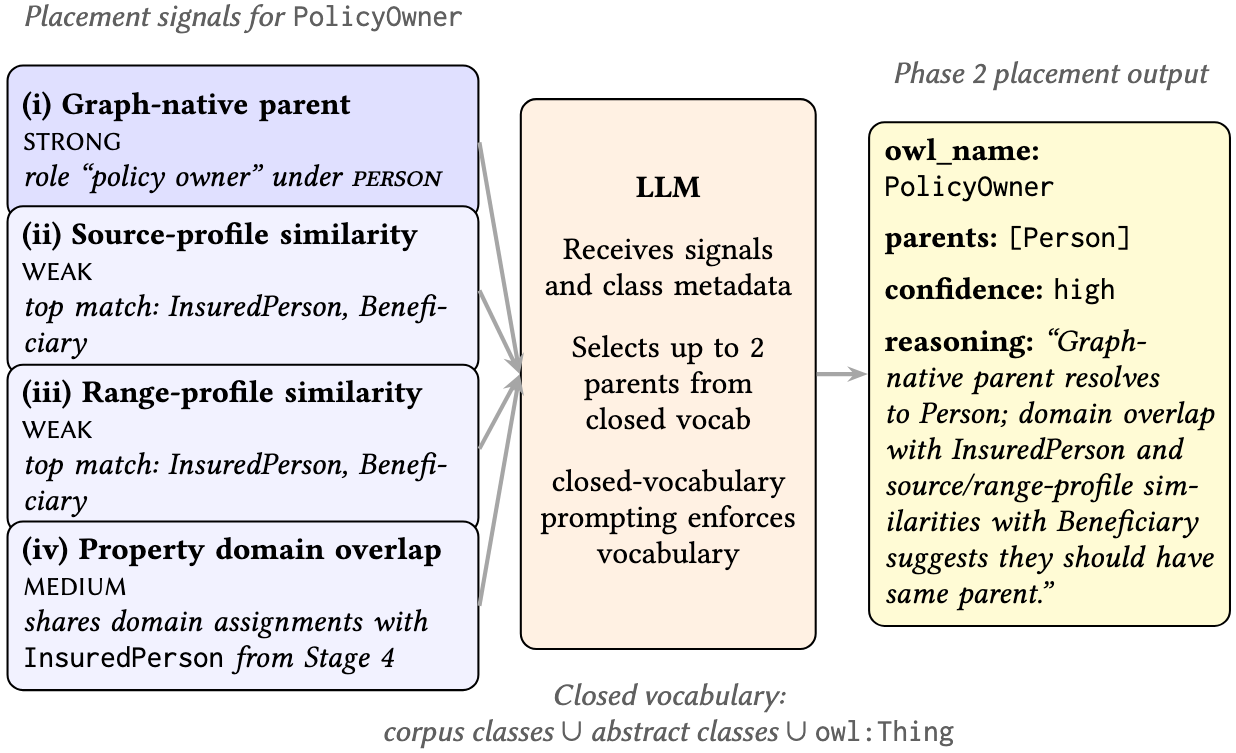}
    \caption{\normalfont\textbf{Hierarchy construction for} \texttt{PolicyOwner}.
    Signal~(i) gives a strong graph-native parent from Stage~3 metadata;
    signals~(ii)--(iii) provide structural evidence from source and range
    edge-type profile similarity; and signal~(iv) identifies sibling relations
    through Stage~4 property-domain overlap. The LLM selects at most two
    parents from a closed vocabulary enforced by closed-vocabulary prompting.}
    \vspace{-5pt}
    \label{fig:stage5-hierarchy}
\end{figure}
}

\paragraph{\textbf{Phase 2: Per-class Placement:}} Each class is placed using four signals.

\textbf{(i) Graph-native parent} resolved in Stage~3 is the strongest signal: a role consistently observed under a given entity type directly encodes a \texttt{subClassOf} hypothesis. For example, the role \emph{policy owner} observed under entity type \textsc{person} strongly indicates $\texttt{PolicyOwner} \sqsubseteq \texttt{Person}$.

\textbf{(ii) Source-profile similarity} is computed from the full 
empirical distribution of source entity types stored per edge label in 
the Stage~2 \texttt{meta} field. For each class, a profile vector is 
constructed over edge labels weighted by frequency as source type. Two 
classes with high cosine similarity over this distribution are 
subsumption candidates: if \texttt{PolicyOwner} and 
\texttt{InsuredPerson} both appear as frequent source types of the same 
set of edge labels, their structural positions in the corpus are similar, 
indicating a likely subsumption or close common-parent relationship.

\textbf{(iii) Range-profile similarity} is constructed symmetrically 
to~(ii): a profile vector is built over edge labels weighted by frequency 
as target type, and two classes with high cosine similarity over this 
distribution are sibling candidates under a common parent, as they occupy 
equivalent structural positions on the receiving end of the same 
relations.

\textbf{(iv) Property domain overlap} is derived independently from 
Stage~4: two classes each assigned as the domain of the same object 
properties share the same relational commitments and are therefore 
sibling or subsumption candidates. Unlike signal~(i), which arises from 
entity role co-occurrence in the graph, this signal is grounded in the 
LLM domain assignments from Stage~4 and can surface relationships between 
classes that never co-occurred as roles in the corpus.

As shown in ~\autoref{fig:stage5-hierarchy}, all four signals, annotated with their strength, are presented to the
LLM in a structured prompt.
The LLM selects one or more parents from a closed vocabulary of corpus-derived
classes, abstract classes, and \texttt{owl:Thing}, enforced via guided
decoding, and returns a confidence rating and brief reasoning string per
assignment.
Low-confidence assignments are flagged automatically for human review.

\vspace{-2pt}
\paragraph{\textbf{Cycle resolution:}} Because placement calls are independent, the resulting hierarchy representation may contain directed cycles, which are invalid in OWL. After all placements are collected, the hierarchy is checked iteratively via depth-first search. 
To handle discovered two-node cyclic relationships such as \texttt{Amount} is a \texttt{subClassOf} \texttt{MonetaryValue} and \texttt{MonetaryValue} is a \texttt{subClassOf} \texttt{Amount}, a dedicated prompt presents the definitions of both classes to the LLM and asks which subsumption direction is semantically more correct. The losing direction is removed and the winner is retained. For a cycle of length more than 2, the minimal cycle path is reconstructed via breadth-first search from the back-edge endpoints.
The full path, class definitions, current parents, and a set of valid replacement anchors are presented to the LLM, which identifies the semantically weakest edge in the cycle and, if removing that edge makes a class parentless,
proposes a replacement parent from the set.
If the LLM call fails or returns an invalid response, a confidence-based
fallback removes the edge involving the lowest-confidence placement and
elevates the affected class to its nearest valid ancestor. Passes repeat until no cycles remain.

\subsection*{Stage 6: Restriction Induction}
\label{subsec:stage6}

This stage converts the restriction-stream clusters from stages 3 and 4 into a final set of OWL TBox axiom patterns. Each cluster carries a \texttt{restriction\_family} label given by Stage~3, encoding the logical
structure of the constraint (e.g., \texttt{cardinality\_constraint},
\texttt{disjointness}, \texttt{eligibility\_condition}), along with typed domain and range assignments from Stage~4.

For each cluster, a candidate is assembled by collecting the same complementary evidence sources from Stage~4 (except S2): the edge-topology distribution over entity types, the raw sentences from fact nodes whose names match the cluster, and the thematic chunk topics from those same facts. These are presented jointly to the LLM so it can cross-check its decisions against corpus-grounded evidence at multiple levels of abstraction: structural, sentential, and thematic. The allowed OWL construct set is derived deterministically from \texttt{restriction\_family}, so the LLM selects from a small pre-constrained set rather than choosing freely among all OWL operators. 
Property names are resolved by token-overlap matching against the closed property vocabulary, since Stage 3 assigns clause-level names to restriction clusters (e.g., \texttt{PolicyLapseAfterGracePeriod}) rather than relational predicates. When Stage 4's range is uninformative, the filler is recovered by scanning the cluster definition for known class names.

The LLM returns a confidence rating alongside each formalization. Low-confidence results are separated, preserving the evidence for human inspection. A final deterministic coherence pass deduplicates identical axiom strings, and resolves contradicting cardinality
assertions on the same property.

\subsection*{Stage 7: TBox Compilation}

This final stage assembles the outputs of all preceding stages into a single, formally valid OWL TBox serialized in Turtle ~\cite{beckett2014rdf} and RDF/XML. The assembly draws from four sources: Stage 5 for the class hierarchy, Stage 3 to back-fill missing definitions and surface variants, Stage 4 for annotated property clusters, and Stage 6 for the induced restriction axioms.
Because all ontological decisions, including naming, domain and range assignment, hierarchical placement, and restriction formalization, have already been made and recorded by the preceding stages, this assembly is entirely deterministic: no LLM call is made and no new decision is introduced. Terms produced with low confidence are included but annotated as such, so that a human auditor can identify and inspect uncertain terms
directly within the TBox without consulting any intermediate pipeline file.

\paragraph{\textbf{Traceability:}}
A central property of the assembled TBox is end-to-end provenance.
Because each stage preserves its decision artifacts, including the
graph-native parent signal and placement confidence from Stage 5,
the domain and range confidence and reasoning string from Stage 4,
the surface variant cluster members and their corpus frequencies from
Stage 3, and the evidence sentences and axiom family from Stage 6,
the assembly has access to a rich chain of evidence for every class,
property, and restriction it emits.
This chain is materialized as human-readable annotations on each
emitted term, making the full decision path inspectable without
consulting any intermediate file.
For example, \texttt{PolicyOwner}'s provenance record shows five surface variants, including \textit{policy owner}, \textit{issuing policy owner}, and \textit{owner of the policy}, merged by Stage~3, its assignment as domain of \texttt{hasPremiumPaymentFrequency} with high confidence from Stage~4, and its placement as a subclass of \texttt{Person} via a strong graph-native parent signal from Stage~5 --- tracing a continuous path from raw policy text to the final OWL declaration. Since ontology induction is inherently uncertain, this provenance makes errors diagnosable and trust in the output evidence-grounded rather than assumed.

\subsection*{ABox Population}

Given the assembled TBox and the per-document typed pools produced by Stage 1, we also design a method for producing a populated OWL ABox for each contract, asserting one named individual per entity and grounding it in the class and property vocabulary established by the TBox. ABox construction proceeds in three steps.

\vspace{-0.5em}
\paragraph{\textbf{Individual Minting:}}
For each entity record, we resolve a set of OWL class names by looking up its type (e.g., \texttt{PERSON}) and each of its atomic roles (e.g., \textit{policy owner}) in indexes built from Stages 3 and 4. When multiple classes are resolved, we select the most derived one in the TBox hierarchy as the canonical class for that individual. Two entity nodes are merged into a single individual if and only if they share the same canonical class and the same surface name, preventing entities with the same name but different roles from being incorrectly collapsed. Each individual is assigned an Internationalized Resource Identifier (IRI) built from its name and contract identifier, and receives \texttt{rdf:type} triples for all resolved classes, a name literal, and any key-value attributes attached to its Stage 1 entity record (e.g., \texttt{dateOfBirth}, \texttt{faceAmount}).

\vspace{-0.5em}
\paragraph{\textbf{Object Property Assertion:}}
We assert object property triples from the edge buckets produced by Stage 1. For \texttt{entity\_entity} edges, we look up the edge label in the property index from Stages 3 and 4 and assert the triple directly between the two individuals. For fact-mediated edges, we first identify the primary subject of each fact from the \texttt{entity\_fact} bucket, then pair it with the target entity from the \texttt{fact\_entity} bucket to form the triple. Edges whose label does not resolve to a known property, or whose source individual's type does not match the property's assigned domain, are discarded.

\vspace{-0.5em}
\paragraph{\textbf{Datatype Property Mining:}}
Fact sentences often contain literal values such as amounts, dates, and durations that cannot be read directly from graph structure. For each fact with a resolvable subject individual, we issue a bounded LLM call with the fact sentence and a closed vocabulary of \texttt{owl:DatatypeProperty} terms drawn from Stages 3 and 4. The model returns property-value pairs, each of which is validated against the vocabulary before being asserted as a literal triple. This follows the same design principle as the TBox pipeline: the graph defines the structure and constrains the vocabulary, while the LLM fills in bounded semantic judgments within those constraints.

\section{Evaluation}

We evaluate our pipeline in the life insurance domain using three complementary evaluation setups, each targeting a different aspect of ontology quality and functional usability. Our evaluation is confined to the life insurance domain; however, since every pipeline stage operates on domain-agnostic graph signals rather than insurance-specific rules or lexicons, we expect the architecture to transfer to other high-stakes, clause-intensive domains, leaving empirical validation to future work.

\subsection{Datasets}

\textbf{Talukder et al.}~\cite{talukder2026towards} study automated ontology generation from life insurance contracts, comparing a direct single-agent LLM baseline against a multi-agent architecture. Their work provides two publicly available synthetic term life insurance contracts: Equivita and Sentinel, along with automatically derived CQs, and evaluates generated ontologies on architectural quality scored by a panel of LLM judges across Extensibility, Redundancy, and Ontology Design Pattern (ODP) Usage, and functional CQ coverage (via SPARQL and RAG-based evaluation). We adopt their contracts, CQ sets, and evaluation setup directly to compare against their baselines under identical conditions.

\textbf{Mridul et al.} ~\cite{mridul2026benchmark} provide a benchmark for gap and overlap analysis as a test of knowledge graph task readiness. It consists of ten expert-verified synthetic life insurance contracts spanning the full range of major product categories, a manually constructed ground-truth domain ontology (TBox) with a fully populated and evidence-annotated ABox where every individual is anchored to a contract identifier, section reference, and verbatim clause text, and 58 structured scenarios each paired with SPARQL queries, contract-level ground-truth outcome labels (\textsc{Covered}, \textsc{Denied}, \textsc{Not\_Applicable}), and clause-level justifications. The scenarios were generated using an LLM-assisted workflow and subsequently curated by a domain expert, with eight expert-authored edge cases probing complex legal nuances. The ground-truth ontology and its SPARQL queries achieve 100\% accuracy by construction, and the task complexity is confirmed by the fact that state-of-the-art LLMs reading the contracts directly achieve only 65--87\% accuracy. We use this benchmark as a primary evaluation substrate and extend it from ten to one hundred contracts for scalability analysis.

{
\setlength{\abovecaptionskip}{2pt}
\setlength{\belowcaptionskip}{-4pt}
\begin{figure}[t]
    \centering
    \includegraphics[width=1\linewidth]{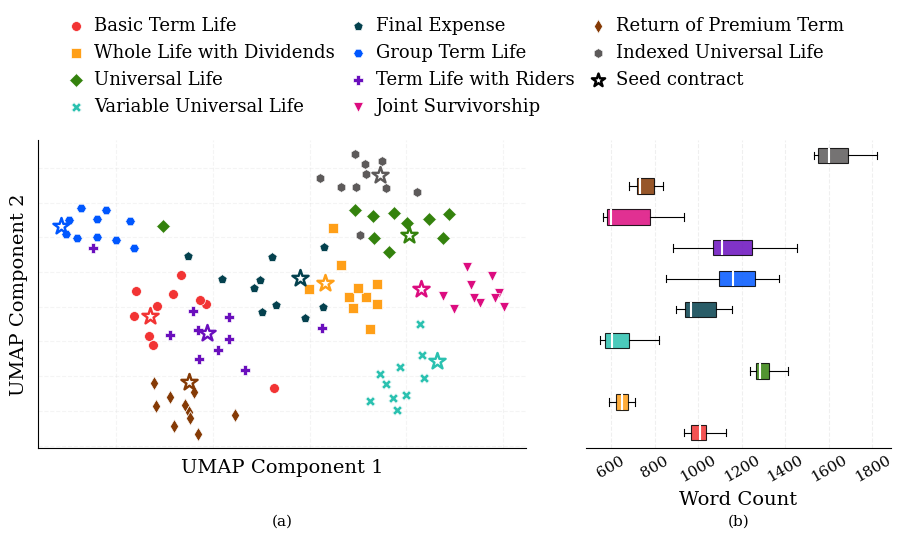}
    \caption{\normalfont(a) \textbf{UMAP~\cite{mcinnes2018umap} projection of the 100-contract corpus.} Contracts naturally separate by product type, yet within each cluster the contracts show substantial within-cluster dispersion, confirming that variants are diverse rather than near-duplicates of the seed contracts. (b) \textbf{Word-count distribution per product type.}}
    \label{fig:contract_diversity}
    \vspace{-0.75em}
\end{figure}
}

\textbf{Extended 100-contract dataset.} We generated an extended 100-contract synthetic corpus using Claude Sonnet 4.6~\cite{anthropic2026claude46}, Gemini-3~\cite{google2025gemini3}, and ChatGPT-5.3~\cite{openai2026chatgpt53}, using the original ten contracts as product-type seeds. ~\autoref{fig:contract_diversity}(a) shows a UMAP~\cite{mcinnes2018umap} projection of the resulting corpus. The contracts separate clearly by product type, and the embedding visualization provides no obvious evidence of collapse to near-duplicates of the seed documents. Within each cluster, the generated variants spread across the embedding space, indicating meaningful structural diversity.  Word-count distributions per product type are shown in ~\autoref{fig:contract_diversity}(b).

\subsection{Architectural Quality and CQ Coverage}

Tables~\ref{tab:cq_coverage} and~\ref{tab:quality_metrics} compare our pipeline against the direct generation and multi-agent baselines from Talukder et al.~\cite{talukder2026towards} on their Equivita and Sentinel contracts.

\begin{table}[t]
\centering\small
\setlength{\tabcolsep}{8pt}
\caption{\normalfont CQ Coverage Metric comparing our pipeline against the direct generation and multi-agent baselines from Talukder et al.~\cite{talukder2026towards}.}
\label{tab:cq_coverage}
\begin{tabular}{llccc}
\toprule
\textbf{Sample} & \textbf{Method} & \textbf{Direct} & \textbf{Multi-Agent} & \textbf{Ours} \\
\midrule
Equivita & SPARQL & 0.6304 & 0.6196 & \textbf{0.8527} \\
         & RAG              & 0.5994 & \textbf{0.7112} & 0.6646 \\
\midrule
Sentinel & SPARQL & 0.4027 & 0.4383 & \textbf{0.7694} \\
         & RAG              & 0.4450 & 0.5510 & \textbf{0.6812} \\
\bottomrule
\end{tabular}
\end{table}

{
\setlength{\abovecaptionskip}{2pt}
\setlength{\belowcaptionskip}{-5pt}
\begin{table}[t]
\centering\small
\setlength{\tabcolsep}{5pt}
\caption{\normalfont Architectural quality metrics comparing our pipeline against the direct generation and
multi-agent baselines from Talukder et al.~\cite{talukder2026towards}. Scores are averaged across 5 LLM judges. Inter-judge reliability measured using \textbf{Krippendorff's $\alpha$}~\cite{hayes2007answering} over the combined judged artifact set is \textbf{0.71} for Extensibility,
\textbf{0.57} for Redundancy, \textbf{0.61} for ODP Usage, and \textbf{0.73} overall.}
\label{tab:quality_metrics}
\begin{tabular}{llccc}
\toprule
\textbf{Dataset} & \textbf{Metric} & \textbf{Direct} & \textbf{Multi-Agent} & \textbf{Ours} \\
\midrule
Equivita & Extensibility  & 1.60 & 3.80          & 3.60 \\
         & Redundancy     & 1.00 & 2.40          & \textbf{2.60} \\
         & ODP Usage      & 1.40 & \textbf{3.80} & 3.60 \\
         & Overall        & 1.33 & \textbf{3.33} & 3.27 \\
\midrule
Sentinel & Extensibility  & 2.80 & 3.80          & \textbf{4.00} \\
         & Redundancy     & 2.00 & 1.80          & \textbf{3.10} \\
         & ODP Usage      & 2.20 & 3.40          & \textbf{3.50} \\
         & Overall        & 2.33 & 3.00          & \textbf{3.53} \\
\bottomrule
\end{tabular}
\vspace{-1.25em}
\end{table}
}

On SPARQL-based CQ coverage (~\autoref{tab:cq_coverage}), our pipeline outperforms both baselines. The RAG-based scores follow the same trend on Sentinel, where we achieve 0.6812 against 0.4450 and 0.5510. On Equivita RAG, our score of 0.6646 is below the multi-agent score of 0.7112. RAG evaluation synthesizes retrieved ontology subgraphs into natural language answers before adjudication, which can favor ontologies whose graph topology and labeling patterns align well with the embedding-based retrieval used in that specific evaluation setup. This is a known sensitivity of RAG-based metrics to surface representation choices, and the result does not reflect a gap in knowledge coverage, as our strong SPARQL score on the same contract confirms.

On architectural quality (~\autoref{tab:quality_metrics}), our pipeline performs on par with or better than the multi-agent baseline and substantially above the direct generation baseline on both contracts. The redundancy improvement is meaningful because it is architecturally the hardest dimension to optimize: it requires global semantic consistency across the ontology, which LLM-based systems struggle with due to attention limitations and the absence of an external concept registry. Our embedding-based clustering stage directly addresses this by deduplicating surface variants before any OWL term is introduced. We note that these scores are produced by LLM judge panels on a coarse 1--5 scale using relatively generic evaluation prompts, so narrow score differences should not be over-interpreted. What the results establish clearly is that our pipeline does not sacrifice structural quality for coverage gains.

\subsection{Ontology Keyphrase Coverage}

We apply the keyphrase coverage methodology from Mridul et al.~\cite{mridul2026benchmark}, which extracts domain keyphrases from contracts using Claude Sonnet 4.6 and matches them against TBox artifacts using three complementary methods: exact/substring matching, fuzzy matching (RapidFuzz\footnote{https://pypi.org/project/RapidFuzz} token-sort ratio $\geq$ 70), and semantic similarity (SentenceTransformer\footnote{https://pypi.org/project/sentence-transformers} cosine similarity $\geq$ 0.70). A keyphrase is considered covered if any method produces a match.

{
\setlength{\abovecaptionskip}{2pt}
\setlength{\belowcaptionskip}{-4pt}
\begin{figure}[t]
    \centering
    \vspace{-3pt}
    \includegraphics[width=1\linewidth]{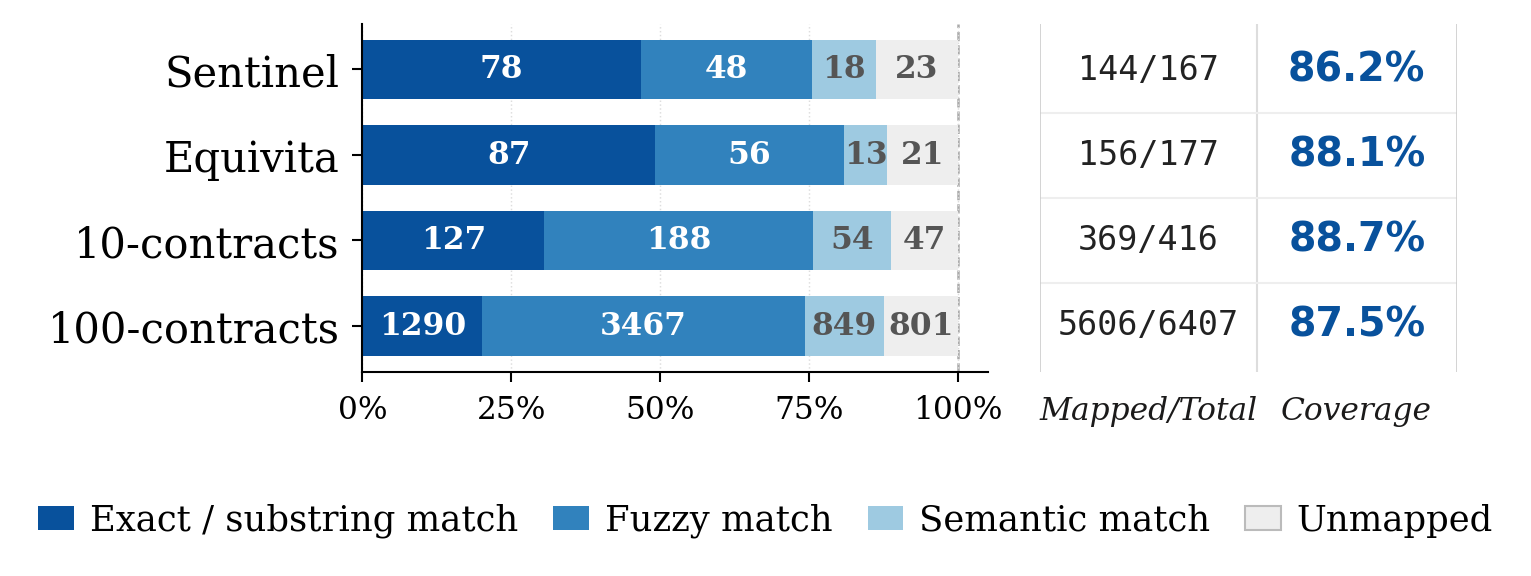}
    \caption{\normalfont Ontology keyphrase coverage analysis across contract sets --- Sentinel and Equivita from Talukder et al.~\cite{talukder2026towards}, 10-contract dataset from Mridul et al.~\cite{mridul2026benchmark}, and our extended 100-contract dataset. For each dataset, keyphrases extracted from the contracts are matched against TBox artifacts using \textbf{substring, fuzzy, and semantic similarity matching.}}
    \label{fig:tbox_coverage}
\end{figure}
}

~\autoref{fig:tbox_coverage} reports results across all four evaluation corpora. Our pipeline achieves strong coverage in all settings: 86.2\% for Sentinel, 88.1\% for Equivita, 88.7\% for the 10-contract dataset, and 87.5\% for the 100-contract dataset. These figures are consistent with the 84.6\% coverage reported by Mridul et al.~\cite{mridul2026benchmark} for their hand-constructed ontology. As the original benchmark paper notes, the small unmapped fraction does not necessarily always indicate ontology gaps. Manual inspection shows that in most cases the relevant concept is represented under a different label or through a richer structure that string-level matching cannot resolve. This pattern holds for our pipeline as well.

\subsection{CQ Coverage on the 10-Contract Dataset}

To evaluate functional usability against the Mridul et al.~\cite{mridul2026benchmark} benchmark, we generated 300 competency question--answer pairs (30 per contract) from the 10-contract dataset using Claude Sonnet 4.6. For each question, we generated a corresponding SPARQL query grounded in our induced ontology's vocabulary and executed it against the ABox produced by our pipeline's ABox construction stage. Query results were evaluated against the reference answers using Qwen3-30B-Instruct~\cite{qwen3technicalreport} as an LLM judge. The evaluation pipeline is illustrated in ~\autoref{fig:cq_overview}(a).

~\autoref{fig:cq_overview}(b) shows per-contract accuracy scores. Our pipeline achieves 91\% overall accuracy, with per-contract scores ranging from 0.73 to 1.00. Strong performance holds even for structurally complex products such as C4 (Variable Universal Life, 0.96) and C10 (Indexed Universal Life, 0.88), demonstrating that the induced ontology and populated ABox are sufficient to answer the large majority of factual CQs derivable from the source contracts.

{
\setlength{\abovecaptionskip}{2pt}
\setlength{\belowcaptionskip}{-6pt}
\begin{figure}[t]
    \centering

    \begin{minipage}{\linewidth}
        \centering
        \includegraphics[width=0.8\linewidth]{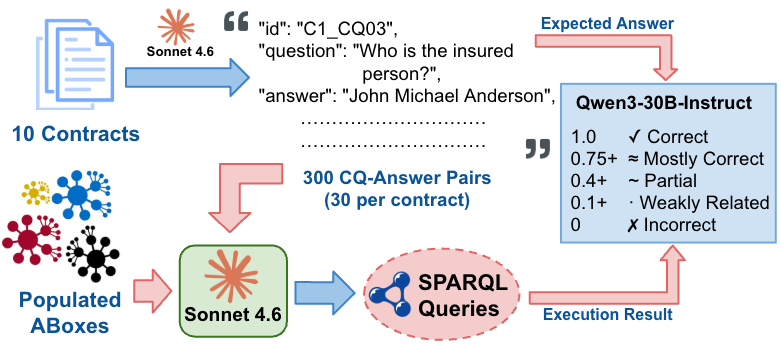}
        \\[4pt]
        \textbf{(a)}
        \label{fig:cq_generation}
    \end{minipage}

    \vspace{0.1cm}

    \begin{minipage}{\linewidth}
        \centering
        \includegraphics[width=0.85\linewidth]{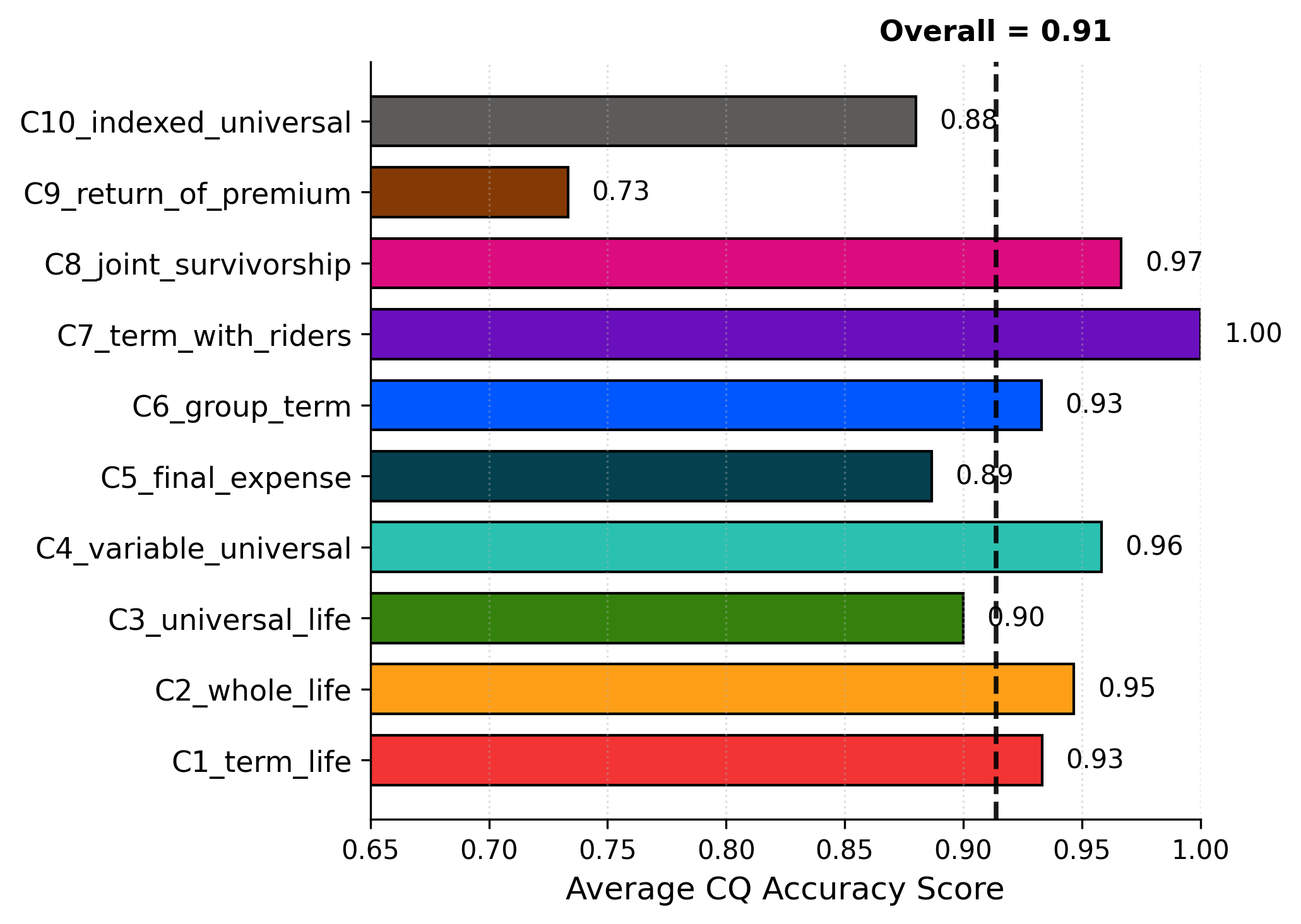}
        \\[4pt]
        \textbf{(b)}
        \label{fig:cq_coverage}
    \end{minipage}

    \caption{\normalfont \textbf{CQ evaluation on the 10-contract dataset~\cite{mridul2026benchmark}.} \textbf{(a) Pipeline:} A total of 300 CQs are generated using Claude Sonnet 4.6, along with SPARQL queries grounded in the populated ABoxes. Query execution results are compared against ground truth answers using Qwen3-30B-Instruct as an LLM judge. \textbf{(b) CQ Coverage Scores:} The generated ontology can answer CQs with high accuracy (91\% overall accuracy).}
    \vspace{-0.5em}
    \label{fig:cq_overview}
\end{figure}
}

\vspace{-4pt}
\subsection{Gap and Overlap Analysis}
We evaluate our generated ontology against the gap and overlap benchmark of Mridul et al.~\cite{mridul2026benchmark}. Given a scenario describing an insurance event with specific conditions, the task is to determine for each contract whether the scenario is \textsc{Covered}, \textsc{Denied}, or \textsc{Not\_Applicable}, with justification traceable to contract clauses. The 58 benchmark scenarios probe genuine structural differences across contracts and are confirmed to be non-trivial: reading contract text directly, Claude Sonnet 4.6 achieves 87.76\%, ChatGPT-5.3 achieves 72.93\%, and Gemini-3 achieves 65.17\%, despite the strong document-level question-answering capabilities of state-of-the-art LLMs.

The benchmark provides ground-truth SPARQL queries written against the ground truth ontology. To evaluate our induced ontology, we rewrote these queries using our ontology's vocabulary via ChatGPT-5.5 ~\cite{openai2026chatgpt55}. ~\autoref{tab:llm-overall-accuracy} reports the results. Our induced ontology achieves 86.38\% accuracy, essentially matching the best-performing LLM baseline (Claude at 87.76\%) and outperforming the others. This is a strong result: an automatically induced ontology approaches the strongest direct-LLM baseline without manual TBox engineering, while providing deterministic reproducibility and explicit provenance that LLM inference often cannot offer. The remaining errors reflect modeling imprecisions that can be localized through the pipeline’s provenance annotations.
% This is a strong result: an automatically induced ontology with no manual TBox engineering matches the accuracy of a state-of-the-art LLM reading contracts directly, while additionally providing deterministic reproducibility and structured evidence trails that LLM inference often cannot offer. The remaining gap from the ground-truth reference (100\%) reflects a small number of modeling imprecisions in the induced TBox, precisely the kind of targeted errors that our pipeline's provenance annotations are designed to help a human auditor identify and fix.

\begin{table}[t]
\centering
\caption{\normalfont \textbf{Evaluation on the gap and overlap analysis benchmark~\cite{mridul2026benchmark}.} The generated ontology outperforms ChatGPT-5.3 and Gemini-3 and performs within 1.38 percentage points of Claude Sonnet 4.6, demonstrating strong task-level readiness.}
\vspace{-1em}
\label{tab:llm-overall-accuracy}
\begin{tabular}{lc}
\toprule
\textbf{Method} & \textbf{Accuracy} \\
\midrule
LLM Based - Claude Sonnet 4.6 & 87.76\% ($509/580$) \\
LLM Based - ChatGPT-5.3 & 72.93\% ($423/580$) \\
LLM Based - Gemini-3 & 65.17\% ($378/580$) \\
Ontology Based - Ground Truth & 100.00\% ($580/580$) \\
\midrule
\textbf{Ontology Based - Ours} & \textbf{86.38\% ($501/580$)} \\
\bottomrule
\end{tabular}
\vspace{-1em}
\end{table}

\subsection{Ontology Growth and Scalability}
~\autoref{fig:growth_curve} tracks cumulative ontology elements as a function of documents ingested for both the 10-contract and 100-contract datasets. Both datasets exhibit the same pattern: rapid initial growth as the pipeline encounters new concepts and relations, followed by a steadily decelerating rate of acquisition as the vocabulary saturates. This is the expected behavior of a well-functioning induction pipeline and confirms that the ontology is converging toward a stable domain vocabulary rather than growing unboundedly. In practice, once an ontology has been induced from a sufficiently large corpus, new documents in the same domain will require only incremental updates. In our pipeline, new documents can be appended to the corpus and the pipeline re-run, with the clustering and deduplication stages naturally absorbing known variants and emitting genuinely novel terms where they exist.

\subsection{Auditable Provenance}
Beyond quantitative metrics, one of the most practically significant properties of our pipeline is its comprehensive provenance architecture. Every ontology term carries a complete decision chain: surface variant cluster members and their corpus frequencies from Stage 3, domain and range confidence scores and LLM reasoning strings from Stage 4, graph-native parent signals and placement confidence from Stage 5, and evidence sentences and axiom families from Stage 6. This chain is materialized as human-readable annotations directly within the TBox, allowing any term to be traced back through every pipeline decision to the original document passages that grounded it.

This transforms the induced ontology from a black-box output into an auditable artifact. Even where the generated ontology is not perfect, every uncertain decision is flagged automatically and its supporting evidence is preserved. A human reviewer can inspect these flagged terms, examine the graph evidence and LLM reasoning behind each decision, and apply targeted corrections with full context. Rather than constructing a domain ontology from scratch, a practitioner can run our pipeline, obtain a high-quality draft with the large majority of terms correct, and focus human effort on reviewing and refining the small uncertain subset. This represents a fundamentally more scalable and tractable approach to ontology engineering.

{
\setlength{\abovecaptionskip}{2pt}
\setlength{\belowcaptionskip}{-8pt}
\begin{figure}[t]
    % \vspace{-1em}
    \centering
    \includegraphics[width=\linewidth]{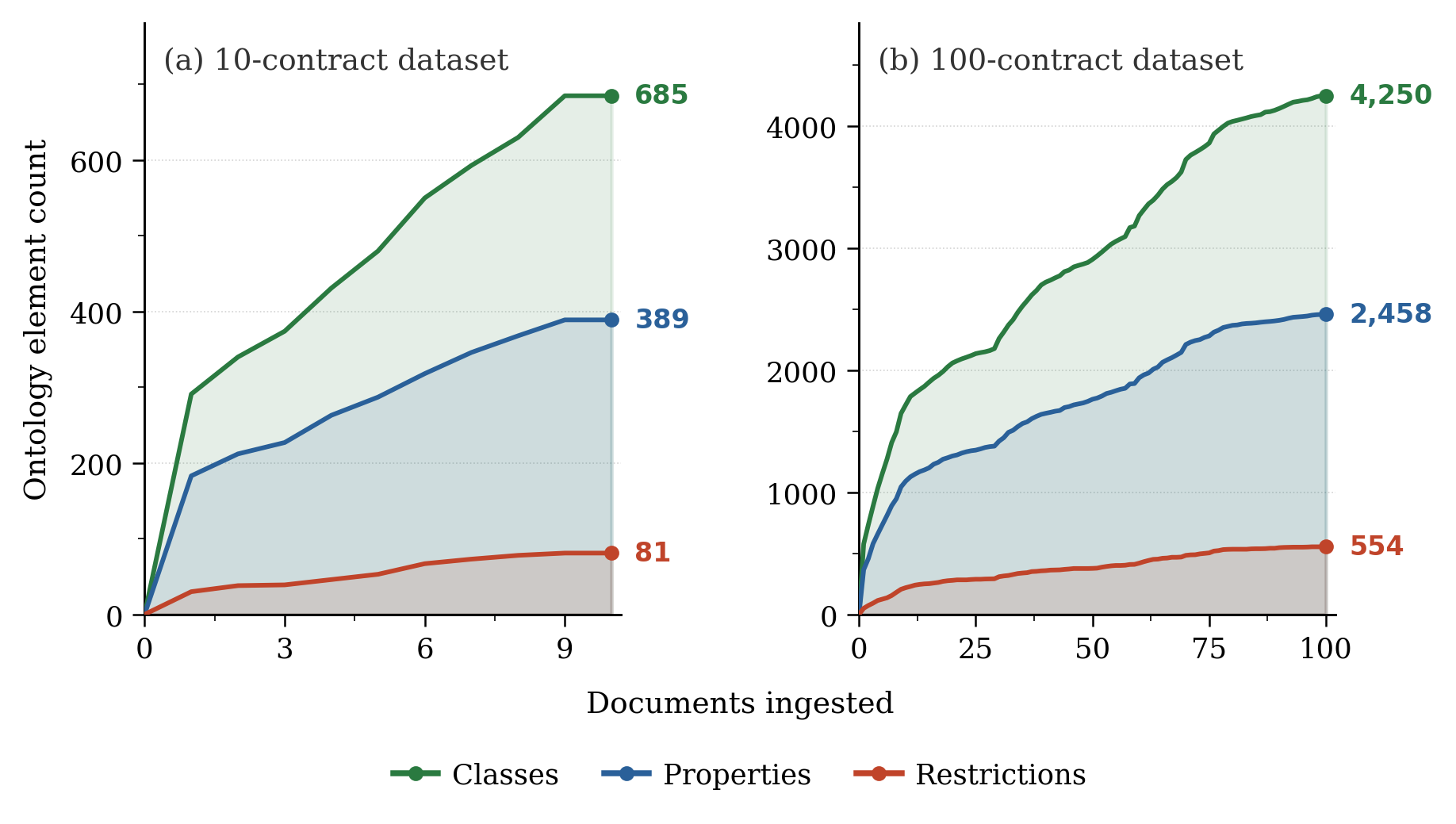}
    \caption{\normalfont \textbf{Ontology growth curves for the 10-contract and 100-contract datasets.} Each curve tracks cumulative ontology elements against documents ingested, showing rapid initial growth followed by decelerating acquisition as the vocabulary saturates.}
    % \vspace{-1.5em}
    \label{fig:growth_curve}
\end{figure}
}

% \vspace{-0.5em}
\section{Conclusion}
We presented a graph-grounded pipeline for ontology induction from 
domain documents that treats LLMs as bounded semantic oracles rather 
than free generators. By first converting documents into UDH and restricting every LLM call to closed-vocabulary 
prompting over corpus-derived evidence, the pipeline produces a complete, 
provenance-annotated OWL TBox and populated ABox without any 
unconstrained generation step. Evaluated on the life insurance domain, 
the induced ontology outperforms direct and multi-agent LLM 
baselines on CQ coverage, matches the best-performing 
LLM on structured gap-and-overlap reasoning without manual TBox 
engineering, and attains keyphrase coverage comparable to a 
hand-constructed reference ontology. Ontology growth analysis confirms 
that the pipeline converges toward a stable domain vocabulary rather than 
growing unboundedly, supporting practical deployment at scale.

Several directions remain open for future work. Because the pipeline's design imposes no insurance-specific structure, its architecture generalizes in principle to other high-stakes document-heavy domains such as legal contracts, clinical guidelines, or financial regulation; empirically validating this transfer is a natural next step. When multiple ontologies are induced independently across organizations, automatically aligning and merging them into a coherent shared vocabulary is a further open challenge that the pipeline's provenance architecture is well-positioned to support. Finally, evaluation of automatically induced ontologies remains a fundamentally hard problem; developing a unified, domain-agnostic benchmark that jointly assesses structural quality, functional coverage, and provenance fidelity would be a valuable contribution to the ontology learning community.

% \section*{Acknowledgments}
% We acknowledge the support from NSF IUCRC CRAFT center research grant (CRAFT Grant \#22022) for this research. The opinions expressed in this publication do not necessarily represent the views of NSF IUCRC CRAFT. We are also grateful for the advice and resources from our CRAFT Industry Board members in shaping this work.

\section*{Acknowledgments}
This research was supported by NSF IUCRC CRAFT Grant \#22022. The views expressed do not necessarily reflect those of NSF IUCRC CRAFT. We thank the CRAFT Industry Board members for their advice and resources.

\section*{GenAI Usage Disclosure}
During the preparation of this work, the authors used Claude Sonnet, ChatGPT, and Gemini to polish and improve the clarity of the text originally drafted by the authors, and for assistance with code segments during implementation. After using these services, the authors reviewed and edited all AI-assisted content as needed and take full responsibility for the publication's content.

\bibliographystyle{ACM-Reference-Format}
\balance
\bibliography{references}

@article{talukder2026towards,
  title={Towards Automated Ontology Generation from Unstructured Text: A Multi-Agent LLM Approach},
  author={Talukder, Abid and Mridul, Maruf Ahmed and Seneviratne, Oshani},
  journal={arXiv preprint arXiv:2604.23090},
  year={2026}
}

@article{mridul2026benchmark,
  title={A Benchmark for Gap and Overlap Analysis as a Test of KG Task Readiness},
  author={Mridul, Maruf Ahmed and Kapa, Rohit and Seneviratne, Oshani},
  journal={arXiv preprint arXiv:2604.10853},
  year={2026}
}

@article{mullner2011modern,
  title={Modern hierarchical, agglomerative clustering algorithms},
  author={M{\"u}llner, Daniel},
  journal={arXiv preprint arXiv:1109.2378},
  year={2011}
}

@article{beckett2014rdf,
  title={RDF 1.1 turtle: Terse RDF triple language},
  author={Beckett, David and Berners-Lee, Tim and Prud’hommeaux, Eric and Carothers, Gavin},
  journal={W3C recommendation},
  volume={25},
  year={2014}
}

@inproceedings{lippolis2025ontology,
  title={Ontology generation using large language models},
  author={Lippolis, Anna Sofia and Saeedizade, Mohammad Javad and Keskis{\""a}rkk{\""a}, Robin and Zuppiroli, Sara and Ceriani, Miguel and Gangemi, Aldo and Blomqvist, Eva and Nuzzolese, Andrea Giovanni},
  booktitle={European Semantic Web Conference},
  pages={321--341},
  year={2025},
  organization={Springer}
}

@incollection{norouzi2025ontology,
  title={Ontology population using LLMs},
  author={Norouzi, Sanaz Saki and Barua, Adrita and Christou, Antrea and Gautam, Nikita and Eells, Andrew and Hitzler, Pascal and Shimizu, Cogan},
  booktitle={Handbook on Neurosymbolic AI and Knowledge Graphs},
  pages={421--438},
  year={2025},
  publisher={IOS Press}
}

@inproceedings{babaei2023llms4ol,
  title={LLMs4OL: Large language models for ontology learning},
  author={Babaei Giglou, Hamed and D’Souza, Jennifer and Auer, S{\""o}ren},
  booktitle={International Semantic Web Conference},
  pages={408--427},
  year={2023},
  organization={Springer}
}

@inproceedings{chowdhury2025automated,
  title={An Automated Framework of Ontology Generation for Abstract Concepts Using LLMs},
  author={Chowdhury, Rafi Rashid and Goto, Takaaki and Tsuchida, Kensei and Kirishima, Tadaaki and Bandi, Ajay},
  booktitle={International Conference on Computers and Their Applications},
  pages={170--180},
  year={2025},
  organization={Springer}
}

@article{abolhasani2024leveraging,
  title={Leveraging LLM for Automated Ontology Extraction and Knowledge Graph Generation},
  author={Abolhasani, Mohammad Sadeq and Pan, Rong},
  journal={CoRR},
  year={2024}
}

@inproceedings{saeedizade2024navigating,
  title={Navigating ontology development with large language models},
  author={Saeedizade, Mohammad Javad and Blomqvist, Eva},
  booktitle={European Semantic Web Conference},
  pages={143--161},
  year={2024},
  organization={Springer}
}

@inproceedings{bakker2024ontology,
  title={Ontology learning from text: an analysis on llm performance},
  author={Bakker, Roos M and Di Scala, Daan L and de Boer, MH},
  booktitle={Proceedings of the 3rd NLP4KGC International Workshop on Natural Language Processing for Knowledge Graph Creation, colocated with Semantics},
  pages={17--19},
  year={2024}
}

@article{lo2024end,
  title={End-to-end ontology learning with large language models},
  author={Lo, Andy and Jiang, Albert Q and Li, Wenda and Jamnik, Mateja},
  journal={Advances in Neural Information Processing Systems},
  volume={37},
  pages={87184--87225},
  year={2024}
}

@article{kommineni2024human,
  title={From human experts to machines: An LLM supported approach to ontology and knowledge graph construction},
  author={Kommineni, Vamsi Krishna and K{\""o}nig-Ries, Birgitta and Samuel, Sheeba},
  journal={CoRR},
  year={2024}
}

@inproceedings{charalambous2022analyzing,
  title={Analyzing coverages of cyber insurance policies using ontology},
  author={Charalambous, Markos and Farao, Aristeidis and Kalantzantonakis, George and Kanakakis, Panagiotis and Salamanos, Nikos and Kotsifakos, Evangelos and Froudakis, Evangellos},
  booktitle={Proceedings of the 17th International Conference on Availability, Reliability and Security},
  pages={1--7},
  year={2022}
}

@inproceedings{ahaggach2023information,
  title={Information extraction and ontology population using car insurance reports},
  author={Ahaggach, Hamid and Abrouk, Lylia and Lebon, Eric},
  booktitle={International Conference on Information Technology-New Generations},
  pages={405--411},
  year={2023},
  organization={Springer}
}

@inproceedings{naqvi12023ontology,
  title={Ontology-Driven Smart Health Insurance},
  author={Naqvi$^1$, Muhammad Raza and Shahzad, Syed Khuram and Iqbal, Muhammad Waseem and Al-Thawadi, Maryam},
  booktitle={Soft Computing Applications: Proceedings of the 9th International Workshop Soft Computing Applications (SOFA 2020)},
  volume={1438},
  pages={51},
  year={2023},
  organization={Springer Nature}
}

@article{el2017towards,
  title={Towards a legal rule-based system grounded on the integration of criminal domain ontology and rules},
  author={El Ghosh, Mirna and Naja, Hala and Abdulrab, Habib and Khalil, Mohamad},
  journal={Procedia computer science},
  volume={112},
  pages={632--642},
  year={2017},
  publisher={Elsevier}
}

@inproceedings{bezerra2013evaluating,
  title={Evaluating ontologies with competency questions},
  author={Bezerra, Camila and Freitas, Fred and Santana, Filipe},
  booktitle={2013 IEEE/WIC/ACM International Joint Conferences on Web Intelligence (WI) and Intelligent Agent Technologies (IAT)},
  volume={3},
  pages={284--285},
  year={2013},
  organization={IEEE}
}

@inproceedings{araujo2016data,
  title={Data-Driven Ontology Evaluation Based on Competency Questions: A Study in the Agricultural Domain},
  author={Ara{\'u}jo, Webert and Lima, Gercina and Pierozzi Jr, Ivo},
  booktitle={Knowledge Organization for a Sustainable World: Challenges and Perspectives for Cultural, Scientific, and Technological Sharing in a Connected Society},
  pages={326--332},
  year={2016},
  organization={Ergon-Verlag}
}

@article{horrocks2004swrl,
  title={SWRL: A semantic web rule language combining OWL and RuleML},
  author={Horrocks, Ian and Patel-Schneider, Peter F and Boley, Harold and Tabet, Said and Grosof, Benjamin and Dean, Mike and others},
  journal={W3C Member submission},
  volume={21},
  number={79},
  pages={1--31},
  year={2004}
}

@inproceedings{amith2022expressing,
  title={Expressing and executing informed consent permissions using SWRL: the all of us use case},
  author={Amith, Muhammad and Harris, Marcelline R and Stansbury, Cooper and Ford, Kathleen and Manion, Frank J and Tao, Cui},
  booktitle={AMIA Annual Symposium Proceedings},
  volume={2021},
  pages={197},
  year={2022}
}

@misc{noy2001ontology,
  title={Ontology development 101: A guide to creating your first ontology},
  author={Noy, Natalya F and McGuinness, Deborah L and others},
  year={2001},
  publisher={Stanford knowledge systems laboratory technical report KSL-01-05 and~…}
}

@article{studer1998knowledge,
  title={Knowledge engineering: Principles and methods},
  author={Studer, Rudi and Benjamins, V Richard and Fensel, Dieter},
  journal={Data \& knowledge engineering},
  volume={25},
  number={1-2},
  pages={161--197},
  year={1998},
  publisher={Elsevier}
}

@article{biemann2005ontology,
  title={Ontology learning from text: A survey of methods},
  author={Biemann, Chris},
  journal={Journal for Language Technology and Computational Linguistics},
  volume={20},
  number={2},
  pages={75--93},
  year={2005}
}

@article{tudorache2020ontology,
  title={Ontology engineering: Current state, challenges, and future directions},
  author={Tudorache, Tania},
  journal={Semantic Web},
  volume={11},
  number={1},
  pages={125--138},
  year={2020},
  publisher={SAGE Publications Sage UK: London, England}
}

@article{du2024short,
  title={A short review for ontology learning: Stride to large language models trend},
  author={Du, Rick and An, Huilong and Wang, Keyu and Liu, Weidong},
  journal={arXiv preprint arXiv:2404.14991},
  year={2024}
}

@article{asim2018survey,
  title={A survey of ontology learning techniques and applications},
  author={Asim, Muhammad Nabeel and Wasim, Muhammad and Khan, Muhammad Usman Ghani and Mahmood, Waqar and Abbasi, Hafiza Mahnoor},
  journal={Database},
  volume={2018},
  pages={bay101},
  year={2018},
  publisher={Oxford University Press}
}

@article{zheng2023judging,
  title={Judging llm-as-a-judge with mt-bench and chatbot arena},
  author={Zheng, Lianmin and Chiang, Wei-Lin and Sheng, Ying and Zhuang, Siyuan and Wu, Zhanghao and Zhuang, Yonghao and Lin, Zi and Li, Zhuohan and Li, Dacheng and Xing, Eric and others},
  journal={Advances in neural information processing systems},
  volume={36},
  pages={46595--46623},
  year={2023}
}

@article{liu2023g,
  title={G-eval: NLG evaluation using gpt-4 with better human alignment},
  author={Liu, Yang and Iter, Dan and Xu, Yichong and Wang, Shuohang and Xu, Ruochen and Zhu, Chenguang},
  journal={arXiv preprint arXiv:2303.16634},
  year={2023}
}

@article{mcinnes2018umap,
  title={Umap: Uniform manifold approximation and projection for dimension reduction},
  author={McInnes, Leland and Healy, John and Melville, James},
  journal={arXiv preprint arXiv:1802.03426},
  year={2018}
}

@article{hayes2007answering,
  title={Answering the call for a standard reliability measure for coding data},
  author={Hayes, Andrew F and Krippendorff, Klaus},
  journal={Communication methods and measures},
  volume={1},
  number={1},
  pages={77--89},
  year={2007},
  publisher={Taylor \& Francis}
}

@article{gruber1993translation,
  title={A translation approach to portable ontology specifications},
  author={Gruber, Thomas R},
  journal={Knowledge acquisition},
  volume={5},
  number={2},
  pages={199--220},
  year={1993},
  publisher={Elsevier}
}

@article{maedche2001ontology,
  title={Ontology learning for the semantic web},
  author={Maedche, Alexander and Staab, Steffen},
  journal={IEEE Intelligent systems},
  volume={16},
  number={2},
  pages={72--79},
  year={2001},
  publisher={IEEE}
}

@book{cimiano2006ontology,
  title={Ontology learning and population from text: algorithms, evaluation and applications},
  author={Cimiano, Philipp},
  year={2006},
  publisher={Springer}
}

@inproceedings{hearst1992automatic,
  title={Automatic acquisition of hyponyms from large text corpora},
  author={Hearst, Marti A},
  booktitle={COLING 1992 volume 2: The 14th international conference on computational linguistics},
  year={1992}
}

@inproceedings{roller2018hearst,
  title={Hearst patterns revisited: Automatic hypernym detection from large text corpora},
  author={Roller, Stephen and Kiela, Douwe and Nickel, Maximilian},
  booktitle={Proceedings of the 56th Annual Meeting of the Association for Computational Linguistics (Volume 2: Short Papers)},
  pages={358--363},
  year={2018}
}

@article{navigli2004learning,
  title={Learning domain ontologies from document warehouses and dedicated web sites},
  author={Navigli, Roberto and Velardi, Paola},
  journal={Computational Linguistics},
  volume={30},
  number={2},
  pages={151--179},
  year={2004},
  publisher={MIT press One Rogers Street, Cambridge, MA 02142-1209, USA journals-info~…}
}

@inproceedings{lourdusamy2019survey,
  title={A survey on methods of ontology learning from text},
  author={Lourdusamy, Ravi and Abraham, Stanislaus},
  booktitle={International Conference on Information, Communication and Computing Technology},
  pages={113--123},
  year={2019},
  organization={Springer}
}

@inproceedings{devlin2019bert,
  title={Bert: Pre-training of deep bidirectional transformers for language understanding},
  author={Devlin, Jacob and Chang, Ming-Wei and Lee, Kenton and Toutanova, Kristina},
  booktitle={Proceedings of the 2019 conference of the North American chapter of the association for computational linguistics: human language technologies, volume 1 (long and short papers)},
  pages={4171--4186},
  year={2019}
}

@inproceedings{reimers2019sentence,
  title={Sentence-bert: Sentence embeddings using siamese bert-networks},
  author={Reimers, Nils and Gurevych, Iryna},
  booktitle={Proceedings of the 2019 conference on empirical methods in natural language processing and the 9th international joint conference on natural language processing (EMNLP-IJCNLP)},
  pages={3982--3992},
  year={2019}
}

@article{willard2023efficient,
  title={Efficient guided generation for large language models},
  author={Willard, Brandon T and Louf, R{\'e}mi},
  journal={arXiv preprint arXiv:2307.09702},
  year={2023}
}

@article{beurer2024guiding,
  title={Guiding llms the right way: Fast, non-invasive constrained generation},
  author={Beurer-Kellner, Luca and Fischer, Marc and Vechev, Martin},
  journal={arXiv preprint arXiv:2403.06988},
  year={2024}
}

@article{geng2025generating,
  title={Generating structured outputs from language models: Benchmark and studies},
  author={Geng, Saibo and Cooper, Hudson and Moskal, Micha{\l} and Jenkins, Samuel and Berman, Julian and Ranchin, Nathan and West, Robert and Horvitz, Eric and Nori, Harsha},
  journal={arXiv e-prints},
  pages={arXiv--2501},
  year={2025}
}

@article{pan2024unifying,
  title={Unifying large language models and knowledge graphs: A roadmap},
  author={Pan, Shirui and Luo, Linhao and Wang, Yufei and Chen, Chen and Wang, Jiapu and Wu, Xindong},
  journal={IEEE Transactions on Knowledge and Data Engineering},
  volume={36},
  number={7},
  pages={3580--3599},
  year={2024},
  publisher={IEEE}
}

@article{zhu2024llms,
  title={Llms for knowledge graph construction and reasoning: Recent capabilities and future opportunities},
  author={Zhu, Yuqi and Wang, Xiaohan and Chen, Jing and Qiao, Shuofei and Ou, Yixin and Yao, Yunzhi and Deng, Shumin and Chen, Huajun and Zhang, Ningyu},
  journal={World Wide Web},
  volume={27},
  number={5},
  pages={58},
  year={2024},
  publisher={Springer}
}

@article{bian2025llm,
  title={LLM-empowered knowledge graph construction: A survey},
  author={Bian, Haonan},
  journal={arXiv preprint arXiv:2510.20345},
  year={2025}
}

@inproceedings{yao2019docred,
  title={DocRED: A large-scale document-level relation extraction dataset},
  author={Yao, Yuan and Ye, Deming and Li, Peng and Han, Xu and Lin, Yankai and Liu, Zhenghao and Liu, Zhiyuan and Huang, Lixin and Zhou, Jie and Sun, Maosong},
  booktitle={Proceedings of the 57th annual meeting of the association for computational linguistics},
  pages={764--777},
  year={2019}
}

@article{velardi2013ontolearn,
  title={Ontolearn reloaded: A graph-based algorithm for taxonomy induction},
  author={Velardi, Paola and Faralli, Stefano and Navigli, Roberto},
  journal={Computational Linguistics},
  volume={39},
  number={3},
  pages={665--707},
  year={2013}
}

@misc{openai2026chatgpt53,
  author = {{OpenAI}},
  title = {ChatGPT (Version 5.3)},
  year = {2026},
  month = {March},
  url = {https://chatgpt.com/},
}

@misc{openai2026chatgpt55,
  author = {{OpenAI}},
  title = {ChatGPT (Version 5.5)},
  year = {2026},
  month = {April},
  url = {https://chatgpt.com/},
}

@misc{google2025gemini3,
  author = {{Google DeepMind}},
  title = {Gemini 3},
  year = {2025},
  month = {December},
  url = {https://gemini.google.com/},
}

@misc{anthropic2026claude46,
  author = {{Anthropic}},
  title = {Claude Sonnet 4.6},
  year = {2026},
  month = {February},
  url = {https://claude.ai/},
}

@misc{qwen3technicalreport,
      title={Qwen3 Technical Report}, 
      author={Qwen Team},
      year={2025},
      eprint={2505.09388},
      archivePrefix={arXiv},
      primaryClass={cs.CL},
      url={https://arxiv.org/abs/2505.09388}, 
}

@misc{udh2026,
  author       = {Talukder, Abid and Mridul, Maruf Ahmed and Seneviratne, Oshani},
  title        = {{Unified Discourse Hypergraph}: Rich graph representation 
                  combining discourse graph and hypergraph},
  year         = {2026},
  publisher    = {GitHub},
  howpublished = {\url{https://github.com/brains-group/unified_discourse_hyper_graph}}
  }

\end{document}